\documentclass[aps,pra,reprint,superscriptaddress]{revtex4-1}

\usepackage[utf8]{inputenc}
\usepackage{amsmath}
\usepackage{graphicx}
\usepackage{multirow}
\usepackage[colorlinks=true, allcolors=blue]{hyperref}
\usepackage{comment}
\usepackage[dvipsnames]{xcolor}
\usepackage[english]{babel}
\usepackage{gensymb}
\renewcommand{\vec}[1]{\boldsymbol{#1}}

\begin{document}
\title{Probing the electromagnetic nonlinearity of vacuum with continuous-wave lasers}
\author{Alexey Arakcheev}
\author{Niv Barkai}
\author{Alexandr Vasilyev}
\author{Osip Schwartz}
\affiliation{Dept. of Physics of Complex Systems, Weizmann Institute of Science, Rehovot, Israel}
\date{\today}

\begin{abstract}\noindent 
In classical electrodynamics, light waves propagating in vacuum do not interact. In quantum physics, however, photon-photon interactions are mediated by virtual particles, giving rise to the electromagnetic nonlinearity of vacuum (EMNV). 
A direct measurement of EMNV would test a long-standing prediction of quantum electrodynamics and constrain new physics models. Despite its fundamental significance and extensive efforts to detect it, free-space EMNV has not yet been directly measured in the laboratory.
Here, we propose a tabletop all-optical measurement of EMNV based on resonantly enhanced four-wave mixing in focusing optical resonators with a circulating power of a few megawatts. As a key experimental step toward this measurement, we demonstrate a resonator reaching a circulating power of 2.5 MW, approaching the parameter range needed to detect EMNV at the level predicted by quantum electrodynamics.
\end{abstract}

\maketitle

\section{Introduction}
\noindent
The coupling of the electromagnetic field to other quantum fields in the physical vacuum gives rise to vacuum polarization and hence to a nonlinear electromagnetic response. In quantum electrodynamics (QED), at photon energies much lower than the electron mass, the electromagnetic nonlinearity of vacuum (EMNV) is described by the Euler-Heisenberg (EH) effective Lagrangian \cite{heisenberg_folgerungen_1936}. 

The quest to observe EMNV is motivated not only by the opportunity to validate a key prediction of QED, but also by the sensitivity of the low-energy interaction constants to physics beyond the Standard Model. Because photon-photon interactions are mediated by virtual particles, they are broadly sensitive to the existence of light particles coupled to the electromagnetic field. In particular, an additional nonlinear term with a distinct polarization dependence could be generated by axion-like particles, which are considered among the main dark matter candidates \cite{bogorad_probing_2019,irastorza_new_2018,evans_virtual_2019, ma_laser-assisted_2025}. 

Photon-photon scattering has been observed in processes involving the strong Coulomb fields of heavy nuclei, such as Delbrück scattering of photons in the nuclear field \cite{schumacher_delbruck_1999}, photon splitting \cite{akhmadaliev_experimental_2002}, and ultraperipheral heavy ion collisions \cite{atlas_collaboration_evidence_2017, atlas_collaboration_observation_2019, hayrapetyan_measurement_2025}. Photon-photon interactions have also been probed in collisions of high-energy electrons with intense laser pulses, where Compton-generated high-energy photons and the laser field produce electron-positron pairs through the nonlinear Breit-Wheeler process \cite{fedotov_advances_2023}.  Additional evidence for EMNV has been provided by optical \cite{mignani_evidence_2017} and X-ray \cite{taverna_polarized_2022, taverna_long_2026} polarimetry of neutron stars, interpreted as signatures of magnetically induced birefringence \cite{lai_ixpe_2023}. %
However, despite this broad range of evidence for photon-photon interactions, free-space EMNV has not yet been directly detected in a terrestrial experiment. 

Such a direct measurement of EMNV is actively pursued across diverse experimental platforms. %
Several long-running experiments (e.g., PVLAS \cite{ejlli_pvlas_2020}, succeeded by VMB@CERN \cite{zavattini_polarimetry_2022}, and BMV \cite{beard_novel_2021}) and new efforts \cite{spector_demonstration_2025} are pursuing the detection of magnetically induced vacuum birefringence via optical polarimetry. These experiments have placed upper bounds on the interaction constants \cite{ejlli_pvlas_2020} but have not yet achieved the sensitivity required to detect the nonlinearity predicted by QED. 

Another experimental effort to measure EMNV is based on four-wave mixing of radio-frequency modes in a high-power superconducting resonator \cite{bogorad_probing_2019, kahn_searching_2022}. While this approach potentially enables probing EMNV at microwave frequencies, it requires detection of the EMNV signal against the nonlinear background generated in the resonator walls \cite{ueki_photon_2024}.

Major X-ray free-electron laser facilities have started studying the interactions of X-ray pulses with high-intensity laser pulses from adjacent ultrahigh-power laser facilities  (HIBEF \cite{ahmadiniaz_towards_2025}, SACLA \cite{inada_probing_2017}, SEL/SHINE \cite{shen_exploring_2018}). However, these existing and planned experiments all face the challenge of isolating a very small EMNV signal from much stronger input beams using polarimetry and spatial separation. Separately, the LUXE \cite{abramowicz_conceptual_2021, borysov_using_2022} collaboration is planning to investigate collisions of high-energy photons in the GeV range with intense laser pulses, as part of its program to investigate strong-field QED both in the perturbative and non-perturbative regimes. 

EMNV is also explored by all-optical methods. The DeLLight experiment aims to measure the deflection of a probe pulse in the nonlinear refractive index gradient induced by an intense pump pulse \cite{robertson_experiment_2021}. The planned  OPAL \cite{rinderknecht_measuring_2025} experiment aims to measure stimulated four-wave mixing between intense laser pulses. A recent proposal calls for a smaller-scale experiment to study laser-induced vacuum birefringence using tightly focused femtosecond laser pulses circulating in high-finesse cavities \cite{luiten_detection_2004, bullis_interferometric_2025}.

Here, we propose a tabletop all-optical approach to measuring EMNV, based on four-wave mixing of continuous-wave (CW) laser beams resonantly enhanced in high-power, high-finesse optical cavities. We calculate the expected signal strength and find that the measurement is feasible with current technology, provided that a near-concentric resonator with a circulating power of a few megawatts can be realized. We then experimentally demonstrate a resonator operating at a record circulating power of 2.5~MW, approaching the power needed for a practical measurement of EMNV at the EH level.

\begin{figure}[th]
    \centering
    \includegraphics[width=8cm]{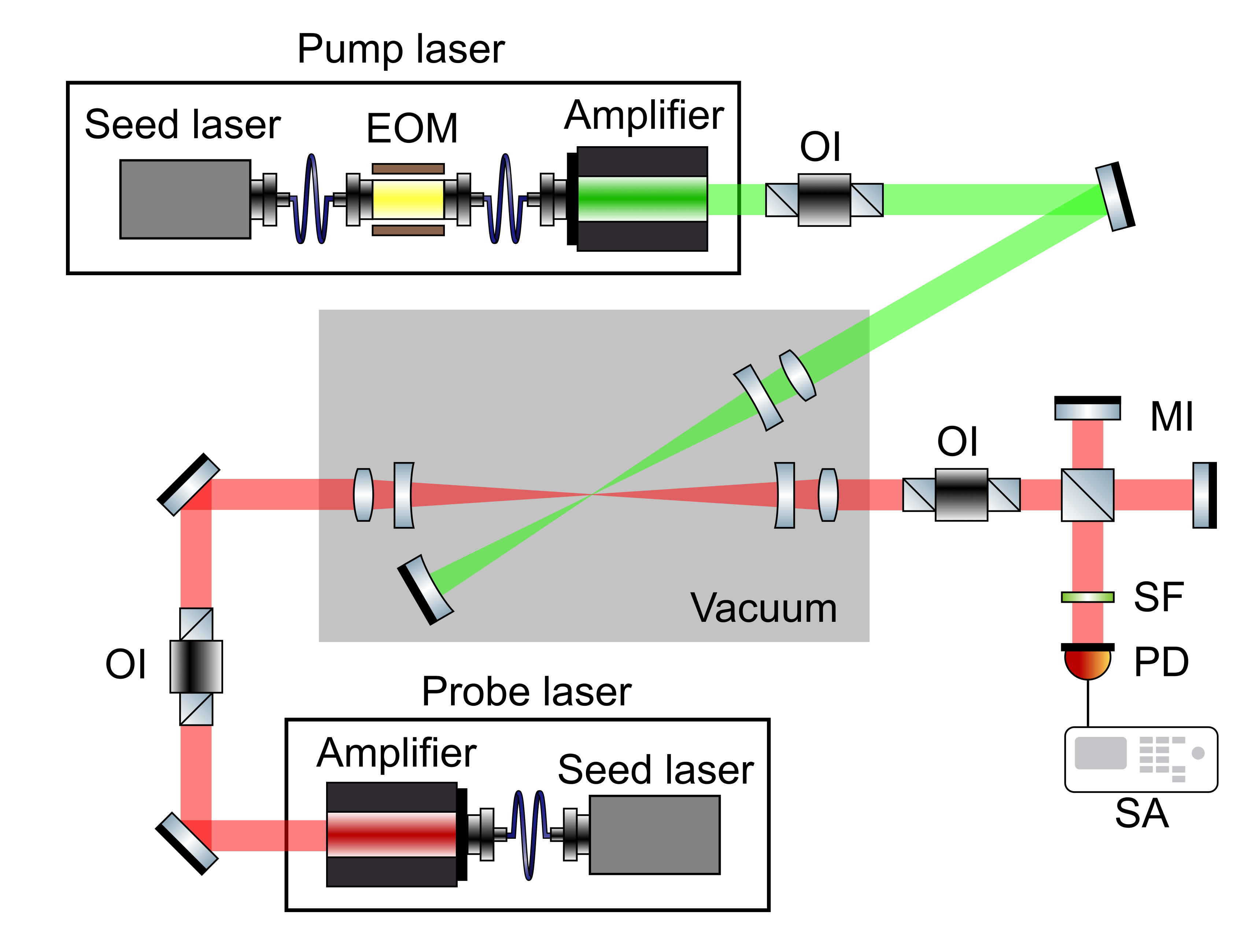}
    \caption{Conceptual schematic of the experiment. Two resonators, the pump and the probe, are suspended in a vacuum chamber. In the pump resonator, two modes are excited, creating a beat note. The EMNV transfers the beat note to the probe beam, inducing phase modulation and creating weak resonant sidebands in the probe resonator. The output beam of the probe resonator passes through a Michelson interferometer, which converts the phase modulation to amplitude modulation and attenuates the carrier, and is directed onto a photodiode for heterodyne detection. Notation:  OI, optical isolator; EOM, electro-optic modulator; MI, Michelson interferometer; SF, spectral filter; PD, photodetector; SA, spectrum analyzer.}
    \label{fig:Scheme}
\end{figure}

\section{Concept of the experiment} \noindent
The experiment is based on a pair of high-power near-concentric Fabry-Perot optical resonators with overlapping focal volumes. One resonator is designated as the pump resonator, and the other as the probe. The two resonators have the same length, and therefore the same free spectral range $\Delta_\text{FSR}$. In the pump resonator, two fundamental transverse modes with different longitudinal mode numbers are excited at frequencies $\omega_1$ and $\omega_2$, so that the frequency difference $\delta\omega = \omega_2-\omega_1$ is an integer multiple of the free spectral range: $\delta\omega =  2\pi \,m  \,\Delta_\text{FSR}$. The two modes create a beat note in the pump resonator, oscillating at the frequency $\delta\omega$, chosen in the range of a few hundred MHz.
Only one fundamental mode of the probe resonator is excited. The probe frequency $\omega_0$ is chosen distinct from that of the pump, so that the probe light can be separated from the pump light using spectral filters.  

The oscillating optical intensity of the pump induces a weak modulation of the probe beam via four-wave mixing in vacuum. Since the two resonators have the same free spectral range, the sidebands created by the modulation at the frequencies $\omega_\pm = \omega_0\pm\delta\omega$ are resonant in the probe resonator. In the time-domain, the amplitude of the modulation builds up over multiple round trips in the resonator, accumulating coherently over the photon storage time. The resonant sideband light is then detected at the output of the probe resonator by heterodyne detection.

A schematic of the experimental apparatus is presented in figure \ref{fig:Scheme}. The two near-concentric resonators with intersecting focal volumes are positioned in a vacuum chamber.  Two laser beams at distinct wavelengths are coupled into the respective resonators. The beam at the output of the probe resonator contains both the original probe frequency and the signal sideband light. The output light is spectrally filtered to remove any scattered pump light. The output beam is then passed through a Michelson interferometer, tuned to fully transmit the signal sidebands while reflecting back most of the carrier light, attenuating it to a level convenient for heterodyne detection. At the same time, the interferometer shifts the relative phase of the sidebands relative to the carrier, converting the phase modulation induced by the nonlinearity into detectable amplitude modulation. 

\section{Calculation of signal strength}\label{Sensitivity to the Euler-Heisenberg nonlinearity} \noindent
The effective nonlinear Lagrangian of the electromagnetic field in vacuum must be Lorentz-invariant. Assuming parity conservation, it must be constructed from the two Lorentz invariants of the electromagnetic field, $\mathcal{F} = \frac{\varepsilon_0}{2} \left(c^2 B^2 - E^2\right)$ and $\mathcal{G} = \varepsilon_0 c \left(\textbf{E} \cdot \textbf{B}\right)$, where $\varepsilon_0$ is the vacuum permittivity and $c$ is the speed of light. Note that throughout this paper, we use the SI system of units.

With only the lowest order nonlinear term taken into account, the Lagrangian generically takes the form:
\begin{equation}
\label{Lagrangian}
\mathcal{L} = -\mathcal{F} + \frac{\xi}{\varepsilon_0} \left(\rho\;  \mathcal{F}^2 + \zeta\;\mathcal{G}^2\right),
\end{equation}
where $\rho$ and $\zeta$ are dimensionless coefficients. A full experimental characterization of four-wave mixing in vacuum thus requires a measurement of both coefficients. 

The QED prediction for the two coefficients was calculated by Euler and Heisenberg  \cite{heisenberg_folgerungen_1936}, yielding $\rho = 4$ and $\zeta =  7$, with 
\begin{equation}
    \xi = \frac{2 \alpha^2 \hbar^3 \varepsilon_0}{45 m_e^4 c^5} = \frac{1}{90 \pi}\frac{\alpha}{E_\text{c}^2}\simeq 1.47\cdot10^{-41}~\mbox{m}^2/\mbox{V}^2, 
\end{equation}
where $\alpha$ is the fine structure constant, $m_e$ is the electron mass, $\hbar$ is the reduced Planck constant,  and $E_\text{c} \simeq 1.32 \cdot 10^{18}$ V/m is the Schwinger field \cite{schwinger_gauge_1951}. 

The sidebands are generated by the electric and magnetic vacuum polarization densities $\textbf{P}$ and $\textbf{M}$:
\begin{equation}
\label{P}
\textbf{P} = \frac{\partial \mathcal{L}_{NL}}{\partial \textbf{E}},\;\textbf{M} = \frac{\partial \mathcal{L}_{NL}}{\partial \textbf{B}},
\end{equation}
where the electric and magnetic fields $\textbf{E}$ and $\textbf{B}$ are, respectively, the sums of the electric and magnetic fields of the pump and probe beams.

The power pumped into the sideband modes can be expressed as:
\begin{multline}
\label{balance}
    \frac{d U_\pm}{d t}  %
    =- \int \textbf{E}_\pm \cdot  \textbf{J}  \;dV =\\
     =- \int \left(\textbf{E}_\pm \cdot \frac{\partial \textbf{P}}{\partial t} -  \textbf{M}\cdot \frac{\partial \textbf{B}_\pm}{\partial t}\right) \;dV,
\end{multline}
where $U_\pm$ is the electromagnetic energy contained in the respective sideband modes, $\textbf{E}_\pm$ and $\textbf{B}_\pm$ denote the electric and magnetic fields of the sidebands, and $\textbf{J}$ is the current induced in vacuum by the pump and probe electromagnetic fields:
\begin{equation}
\label{J}
\textbf{J} = \frac{\partial \textbf{P}}{\partial t} + \left( \nabla \times \textbf{M} \right).
\end{equation}
In the steady state, the pumping of the sidebands by the four-wave-mixing process is balanced by the resonator loss: 
\begin{equation}
\label{balance}
    - \int \textbf{E}_\pm \cdot \textbf{J} \; dV = 2\Gamma U_\pm,
\end{equation}
where $\Gamma = \pi \, \Delta_{FSR}/F$ is the linewidth of the resonator  and $F$ is the finesse of the resonator. 

To find the sideband amplitude explicitly, we express the electric and magnetic fields of each mode in the resonator as:
\begin{align}
    \label{E_mode}
    \textbf{E}_n = \Re \left[ a_n e^{-i \omega_n t} \vec{\mathcal{E}}_n(\textbf{r})\right],\\ \textbf{B}_n = \Re \left[ a_n e^{-i \omega_n t} \vec{\mathcal{B}}_n(\textbf{r})\right],
\end{align}
where $a_n$ is the amplitude, $\vec{\mathcal{E}}(\textbf{r})$ and $\vec{\mathcal{B}}(\textbf{r})$ are the spatial distributions of the electric and magnetic fields. The index $n$ is used to enumerate the modes as follows: $n=0$ denotes the initial probe mode, $n=\pm$ denotes, respectively, the positive and negative-offset signal sidebands, and $n=1,2$  denotes the two excited modes of the pump resonator. The electromagnetic energy stored in the mode then takes the form 
\begin{equation}
    U_n = \frac{\varepsilon_0 \left| a_n^2\right|}{2} \int \left|{\mathcal{E}}_n^2\right| dV.
\end{equation}
Substituting equation (\ref{E_mode}) into the energy balance condition (\ref{balance}), we obtain the sideband amplitude:
\begin{multline}   
\label{a0}
    a_\pm = - \frac{1}{2 \Gamma \varepsilon_0 }  \frac{\int \vec{\mathcal{E}}^*_\pm \cdot \textbf{J}_{\omega \pm}\; dV 
}{\int\left|\mathcal{E}_\pm\right|^2  \;dV } = \\%
=\frac{i \omega_\pm}{2 \Gamma \varepsilon_0 }  \frac{\int \left(\vec{\mathcal{E}}^*_\pm \cdot \textbf{P}_{\omega \pm} + \vec{\mathcal{B}}^*_\pm \cdot \textbf{M}_{\omega \pm} \right)\; dV 
}{\int \left|\mathcal{E}_\pm\right|^2  \;dV },
\end{multline}
where  $\textbf{J}_{\omega\pm}$, $\textbf{P}_{\omega \pm}$, $\textbf{M}_{\omega \pm}$  are the complex amplitudes of the components of $\textbf{J}$, $\textbf{P}$, $\textbf{M}$ oscillating at the respective frequencies of the sideband modes.

To evaluate the sideband amplitude in the proposed experiment, we use the paraxial (Gaussian)  approximation of the fundamental mode shapes in both the pump and probe resonators \cite{siegman_lasers_1986}:
\begin{multline}
\label{GM}
    \vec{\mathcal{E}}_n(\textbf{r}) = \vec{\epsilon}_n \frac{w_{n}}{W_n(z_n)} e^{-{r_{\perp n}^2}/{W_n^2(z_n)}} \times \\ \times \sin\left(k_n z_n + k_n \frac{z_n^2}{2 R_n(z_n)} - \psi_n(z_n) + \frac{\pi}{2} \, l_n\right),
\end{multline}
where  $\vec{\epsilon}_n$ is the polarization unit vector, $k_n=\omega_n/c $, $w_{n}$ is the waist of the mode, $\mathbf{\hat{z}}_n$ is the unit vector along the axis of the corresponding resonator (chosen consistently for all modes in each resonator), $z_n = (\mathbf{\hat{z}}_n \cdot \textbf{r})$ is the coordinate along the optical axis of the resonator, 
\begin{align*}
    \textbf{r}_{\!\perp n} &= \textbf{r} - z_n \mathbf{\hat{z}}_n,\\
    W_n(z_n) &= w_{n} \sqrt{1+\left(\frac{2 z_n}{k_n w_{n}^{2}}\right)^2},\\
    R_n(z_n) &= z_n \left(1 + \left(\frac{k_n w_{n}^{2}}{2 z_n}\right)^2\right),
 \end{align*}
$\psi_n$ is the Gouy phase given by $$\psi_n(z_n) = \arctan\left(\frac{2 z_n}{k_n w_{n}^{2}}\right),$$ and $l_n$ is the longitudinal number of the mode, with even and odd values of $l$ corresponding, respectively, to an electric field node or antinode in the center of the resonator.

Substituting the mode shape (\ref{GM}) into the general formula for the mode amplitude (\ref{a0}), we obtain the following expression for the sideband amplitudes:
\begin{equation}    \label{a1}
a_+ = i\,\chi\,a_0\, a_{1}^*   a_{2}, \quad 
a_\text{-} = i \chi \, a_0\, a_{1}   a_{2}^*, 
\end{equation}
where the constant $\chi$ is:
\begin{equation} \label{chi}
\chi = {\sqrt{\frac{\pi}{2}}}\; \xi \; \frac{K\, F}{\left|\sin\theta\right|} \; \frac{     \; w^{2}_{1} }{ \lambda_0 \sqrt{w_{0}^{2} + w_{1}^{2}} },
\end{equation}%
where $\theta$ is the angle between the axes of the probe and the pump resonators, and $K$ is a dimensionless coefficient depending only on $\theta$, the configuration of polarizations, and the spatial symmetry of the modes.
The form of equation (\ref{a1}) shows that the sideband amplitudes $a_\pm$ are out of phase with the probe field, as should be the case for a purely dispersive interaction.

According to eq. (\ref{chi}), the sideband amplitudes scale inversely with $\sin{\theta}$, while $K$ is nearly constant  at small $\theta$, as shown below. This suggests that the angle $\theta$ should be chosen as small as possible to maximize the signal. However, it cannot be arbitrarily small: to avoid cross-scattering between the resonators, their axes must be separated by an angle at least several times larger than the numerical apertures of the modes. To keep track of this constraint, we introduce the separation factor $s$:
\begin{equation}
\label{s}
s = \frac{\pi}{\lambda_0} \left|\sin\theta \right| \sqrt{\left( w_{0}^{2} + w_{1}^{2}\right)/2}.
\end{equation}

It is also convenient to express the mode amplitude in terms of circulating power in the modes of the pump resonator:
\begin{equation}
\label{CircPower}
P_{n} = \frac{\pi}{16} c \varepsilon_0 w_{n}^{2} |a^2_n| .
\end{equation}
Assuming equal power distribution between the two pump modes $P_1 = P_2 = P_{\text{pump}}/2$, we finally obtain the following expression for the amplitudes of the sideband fields:
\begin{equation}
\label{a2}
\left|\frac{a_\pm}{a_0}\right| = 4  \sqrt{\pi} \;\frac{\xi}{c \varepsilon_0}  \; \frac{P_{\text{pump}}}{ \lambda_0^2} \;\frac{\left| K \right| \,F}{s}
\end{equation}

In practice, it may be easier to excite three equidistant modes in the pump resonator than just two, generating two resonant sidebands at the frequencies  $\omega_{1,3} = \omega_2 \pm \delta\omega$. The maximum sideband generation efficiency in the probe resonator is achieved at the following distribution of power between the pump modes: $P_1 = P_3 = P_{\text{pump}}/4$ and $P_2 = P_{\text{pump}}/2$. In this case, the amplitudes $a_\pm$ increase by a factor of $\sqrt{2}$ relative to eq.~(\ref{a2}).

\subsection{Signal dependence on mode configuration} \noindent
Substituting the mode shape (\ref{GM}) into eq. (\ref{a0}), we find the dimensionless coefficient $K$ for various configurations of the polarization and symmetry of the modes. The dimensionless coefficient $K$ comprises the contributions of the two nonlinear terms in the nonlinear Lagrangian: 
\begin{equation}
\label{K}
K = \rho \, K_{\!\mathcal{F}^2} + \zeta \, K_{\mathcal{G}^2}.
\end{equation}
As shown below, measuring the nonlinear signal for distinct mode configurations enables the determination of both coefficients.

We note that the coefficient $K$ depends on the parity of $m$, where the modulation frequency is $\delta \omega = 2 \pi \, m \,\Delta_\text{FSR}$. For even values of $m$, the two pump modes are spatially in phase in the focal region, and so are the probe and the sideband modes. Conversely, for odd $m$, the two pump modes are spatially $90\degree\!$ out of phase with each other, and so are the signal sidebands with respect to the probe. The difference in spatial overlap leads to distinct polarization dependencies in the odd-$m$ and even-$m$ cases.

To simplify the calculation, we used the following assumptions: $\delta\omega \ll \omega_0$ and $s\gg 1$. These analytical results were validated by numerical calculations using the full Gaussian mode shape, which showed a relative error of less than $0.01$ for the realistic experimental parameters used below.

Four-wave mixing in vacuum generates both cross-phase modulation (XPM) and laser-induced birefringence (BR). Below, we find explicit formulas for the coefficient $K$ both for XPM and BR configurations, in terms of the polarizations of the modes $\vec{\epsilon}_n$ and the unit vectors in the direction of the modes' magnetic fields, defined as $\vec{\beta}_n = \hat{\vec{z}}_n \times \vec{\epsilon}_n$. 

To find the XPM coefficients for arbitrary linear polarizations of the pump and probe fields, we set the polarization of the sideband field equal to that of the probe: $\vec{\epsilon}_\pm = \vec{\epsilon}_0$, $\vec{\beta}_\pm = \vec{\beta}_0$.  Equation (\ref{a0}) then yields the following coefficients:
\begin{align}    
K_{\mathcal{F}^2} &= \left(\vec{\epsilon }_1 \vec{\epsilon}_0\right) \left(\vec{\epsilon }_2 \vec{\epsilon}_0\right) + 
\left(\vec{\beta }_1 \vec{\beta}_0\right) 
\left(\vec{\beta }_2 \vec{\beta}_0\right)  &\text{for even } m & \nonumber\\
K_{\mathcal{F}^2}& = 
-\left(\vec{\epsilon }_1 \vec{\epsilon}_0\right) 
\left(\vec{\beta }_2 \vec{\beta}_0\right) - 
\left(\vec{\beta }_1 \vec{\beta}_0\right) 
\left(\vec{\epsilon }_2 \vec{\epsilon}_0\right) \quad &\text{for odd } m& \nonumber\\
K_{\mathcal{G}^2} &= \left(\vec{\epsilon }_1 \vec{\beta}_0\right) \left(\vec{\epsilon }_2 \vec{\beta}_0\right) + 
\left(\vec{\beta }_1 \vec{\epsilon}_0\right) \left(\vec{\beta }_2 \vec{\epsilon}_0\right) \quad &\text{for even } m & \nonumber\\
K_{\mathcal{G}^2}& = 
\left(\vec{\epsilon }_1 \vec{\beta}_0\right) 
\left(\vec{\beta }_2 \vec{\epsilon}_0\right) + 
\left(\vec{\beta }_1 \vec{\epsilon}_0\right) 
\left(\vec{\epsilon }_2 \vec{\beta}_0\right) \quad &\text{for odd } m& \label{G2odd}
\end{align}
The above equations can be straightforwardly generalized to elliptic polarizations by substituting complex polarization vectors instead of real ones, with complex conjugate polarization vectors standing for emitted waves, and setting $\vec{\epsilon}_\pm = \vec{\epsilon}_0^*$, $\vec{\beta}_\pm = \vec{\beta}_0^*$.

Likewise, to find the coefficients for birefringence for arbitrary linear polarizations of the input fields, we set $\vec{\epsilon}_\pm = \vec{\beta}_0$, $\vec{\beta}_\pm = -\vec{\epsilon}_0$. In this case, we obtain:
\begin{widetext}

\begin{align}    
K_{\mathcal{F}^2}^\text{BR} &=\frac{1}{2} \Big( 
\left(\vec{\epsilon }_1 \vec{\epsilon}_0\right)
\left(\vec{\epsilon }_2 \vec{\beta}_0\right) + 
\left(\vec{\epsilon }_1 \vec{\beta}_0\right) 
\left(\vec{\epsilon }_2 \vec{\epsilon}_0\right) 
-\left(\vec{\beta }_1 \vec{\beta}_0\right)
\left(\vec{\beta }_2 \vec{\epsilon}_0\right)  
-\left(\vec{\beta }_1 \vec{\epsilon}_0\right) 
\left(\vec{\beta }_2 \vec{\beta}_0\right) 
\Big) \quad &\text{for even } m &\nonumber\\
K_{\mathcal{F}^2}^\text{BR} & = \frac{1}{2} \Big(
\left(\vec{\epsilon }_1 \vec{\epsilon}_0\right) 
\left(\vec{\beta }_2 \vec{\epsilon}_0\right) 
- 
\left(\vec{\epsilon }_1 \vec{\beta}_0\right)
\left(\vec{\beta }_2 \vec{\beta}_0\right) 
-
\left(\vec{\beta }_1 \vec{\beta}_0\right)
\left(\vec{\epsilon }_2\vec{\beta}_0\right)
+
\left(\vec{\beta }_1 \vec{\epsilon}_0\right)
\left(\vec{\epsilon }_2\vec{\epsilon}_0\right)
\Big)\quad &\text{for odd } m&\nonumber\\
K_{\mathcal{G}^2}^\text{BR} &=\frac{1}{2} \Big(
-\left(\vec{\epsilon }_1 \vec{\beta}_0\right)
\left(\vec{\epsilon }_2 \vec{\epsilon}_0\right) 
-\left(\vec{\epsilon }_1 \vec{\epsilon}_0\right)
\left(\vec{\epsilon }_2 \vec{\beta}_0\right) 
+\left(\vec{\beta }_1 \vec{\beta}_0\right)
\left(\vec{\beta }_2 \vec{\epsilon}_0\right) 
+\left(\vec{\beta }_1 \vec{\epsilon}_0\right)
\left(\vec{\beta }_2 \vec{\beta}_0\right) 
\Big)\quad &\text{for even } m &\nonumber\\%
K_{\mathcal{G}^2}^\text{BR} & = \frac{1}{2} \Big(
\left(\vec{\epsilon }_1 \vec{\beta}_0\right)
\left(\vec{\beta }_2 \vec{\beta}_0\right) 
-\left(\vec{\epsilon }_1 \vec{\epsilon}_0\right)
\left(\vec{\beta }_2 \vec{\epsilon}_0\right) 
-\left(\vec{\beta }_1 \vec{\epsilon}_0\right)
\left(\vec{\epsilon }_2 \vec{\epsilon}_0\right) 
+\left(\vec{\beta }_1 \vec{\beta}_0\right)
\left(\vec{\epsilon }_2 \vec{\beta}_0\right) 
\Big) \quad &\text{for odd } m%
\end{align}

\end{widetext}

For example, in the case of resonators intersecting at a small angle, with the two pump modes having the same linear polarization, and the probe beam polarized orthogonally to the pump, the XPM coefficients become $K_{\mathcal{G}^2} = 2$ and $K_{\mathcal{F}^2} = 0$,  for both odd and even $m$. Thus, detecting the XPM signal in this configuration would provide a measurement of the coefficient $\zeta$.  This configuration produces the highest possible coefficient for the Euler-Heisenberg nonlinearity, $K_{EH}=14$. The coefficient $\rho$ can be measured independently by detecting XPM in the configuration where both pump modes and the probe have the same linear polarization. In this case, $K_{\mathcal{F}^2} = 2$ and $K_{\mathcal{G}^2} = 0$, resulting in $K_{EH}=8$. In both examples above, the birefringence coefficients vanish.

At a fixed separation factor $s$, eq. (\ref{a2}) shows that the sideband amplitudes $a_\pm$ depend on the angle $\theta$ between the resonators only through the coefficients $K$, which are nearly constant at small $\theta$. Thus, at a given circulating power, the expected signal  is almost independent of the numerical aperture of the resonators. This observation is easily explained: the nonlinear-optically induced phase shift of the probe beam is proportional to the pump intensity at the resonator focus and the interaction length. The former scales as $\text{NA}^2$, while the latter scales as $\text{NA}^{\!-2}$, leading to a NA-independent signal. Thus, the experiment can be carried out with resonators of moderate NA.

\subsection{Shot noise-limited SNR}\label{Signal detection strategy}\noindent
The phase-modulated probe beam exits the resonator through the output mirror. The beam then passes through a dispersive element, such as a Michelson interferometer, that shifts the sideband phase relative to the carrier, converting the phase modulation into amplitude modulation. The power of the output beam takes the following form:
\begin{equation}
\label{PhotonFlux}
P_\text{out} = P_0 + 4 \sqrt{P_0 P_\pm} \cos\left(\delta\omega \,t\right),
\end{equation}
where $P_0 = P_\text{probe} T_\text{probe} $ is the power of the probe beam transmitted through the resonator, and $P_\pm = P_0 \left|{a_\pm}/{a_0}\right|^2$ is the power in the sidebands. A photodiode with a quantum efficiency $\eta$ converts the light into an electric current:
\begin{equation}
\label{Current}
I = \frac{\eta\; e}{\hbar\, \omega_0} \left(P_0 + 4  \sqrt{P_0 P_\pm} \cos\left(\delta\omega t\right)\right).
\end{equation}
The standard deviation of shot noise is $e \sqrt{2 \frac{\eta }{\hbar \omega_0}P_0 \Delta f}$, where $\Delta f = {1}/{(2 \tau)}$ is the bandwidth of measurement of duration $\tau$ \cite{hobbs_building_2022}. 
The shot-noise-limited SNR then takes the form:
\begin{equation}
\label{SNR}
\mbox{SNR} = 2 \sqrt{2} \sqrt{\frac{\eta }{\hbar \omega_0}P_\pm \,\tau}.
\end{equation}
As expected, the signal-to-noise ratio does not depend on the amplitude of the local oscillator. Thus, the amplitude of the carrier can be reduced by a spectral filter to a comfortable level for measurement by a photodiode. Finally, for the integration time required to reach a desired SNR $Q$, we obtain:

\begin{multline}
\label{tau}
\tau = 
\frac{1}{128 \pi }\frac{\hbar\,  c^3 \varepsilon_0^2} {\xi^2 \eta}
\frac{Q^2 s^2  }{K_{EH}^2 }\;
\frac{ \lambda_0^3}{   P_{\text{pump}}^2  P_\text{probe} F } \times \\
\times \frac{  L_\text{probe}+T_\text{probe}}{ T_\text{probe} },
\end{multline}
where $T$ and $L$ are, respectively, the transmission and loss coefficients of the resonator mirrors.

We now substitute the following realistic parameters:
\begin{itemize}
    \item Resonator length: $l= 1$ m; free spectral range: $\Delta_\text{FSR} = 150$ MHz;
    \item Wavelengths of light used in the pump and probe resonators:  $\lambda_1 = 1030$ nm, $\lambda_0 = 1064$ nm;
    \item Modulation frequency: $\delta \omega  = 2 \pi \cdot 300$ MHz, corresponding to $ m = 2 $;
    \item Numerical aperture of the pump and probe resonators: $\mbox{NA} = 5 \times 10^{-3}$; mode waist: $w_0 =66$ $\mu$m; 
    \item Angle between the axes of the pump and probe resonators: $\theta = 0.04$ rad $\simeq 2.3^\circ$; separation factor:  $s = 7.9$;
    \item Power transmission and loss coefficients of the resonator mirrors: $T = L = 5 \cdot 10^{-6}$;
    \item Input laser power: $P_{input} = 100 $ W, corresponding to a circulating power of $5$ MW;
    \item   Photodiode quantum efficiency: $\eta = 0.8$.
\end{itemize}

The signal strength in this experiment can be expressed as the photon flux observed in the signal sidebands, which equals $5.1 \cdot 10^{-6}$ photons/s (in the two sidebands combined) with the parameters listed above. 
At this signal level, to achieve an SNR of $Q=5$, corresponding to a 5-sigma detection significance, at $K = 14$, requires approximately 18 days of integration time, which is feasible in a single experimental run. Since optical fields of similar magnitude have been successfully measured by heterodyne detection in the shot-noise-limited regime \cite{bush_coherent_2019}, we conclude that the EMNV signal strength in this scheme is sufficient for measurement with current technology, so long as the enhancement resonators can reach megawatt-scale power.

\section{Experimental demonstration of a multi-megawatt resonator}
\noindent
Performing an EMNV measurement with CW lasers requires optical resonators with circulating powers of a few megawatts. To the best of our knowledge, the highest average circulating power reported to date was $710$ kW \cite{lu_710_2024}. We therefore set out to determine whether multi-megawatt circulating power is attainable.

\begin{figure}
    \centering
    \includegraphics[width=8cm]{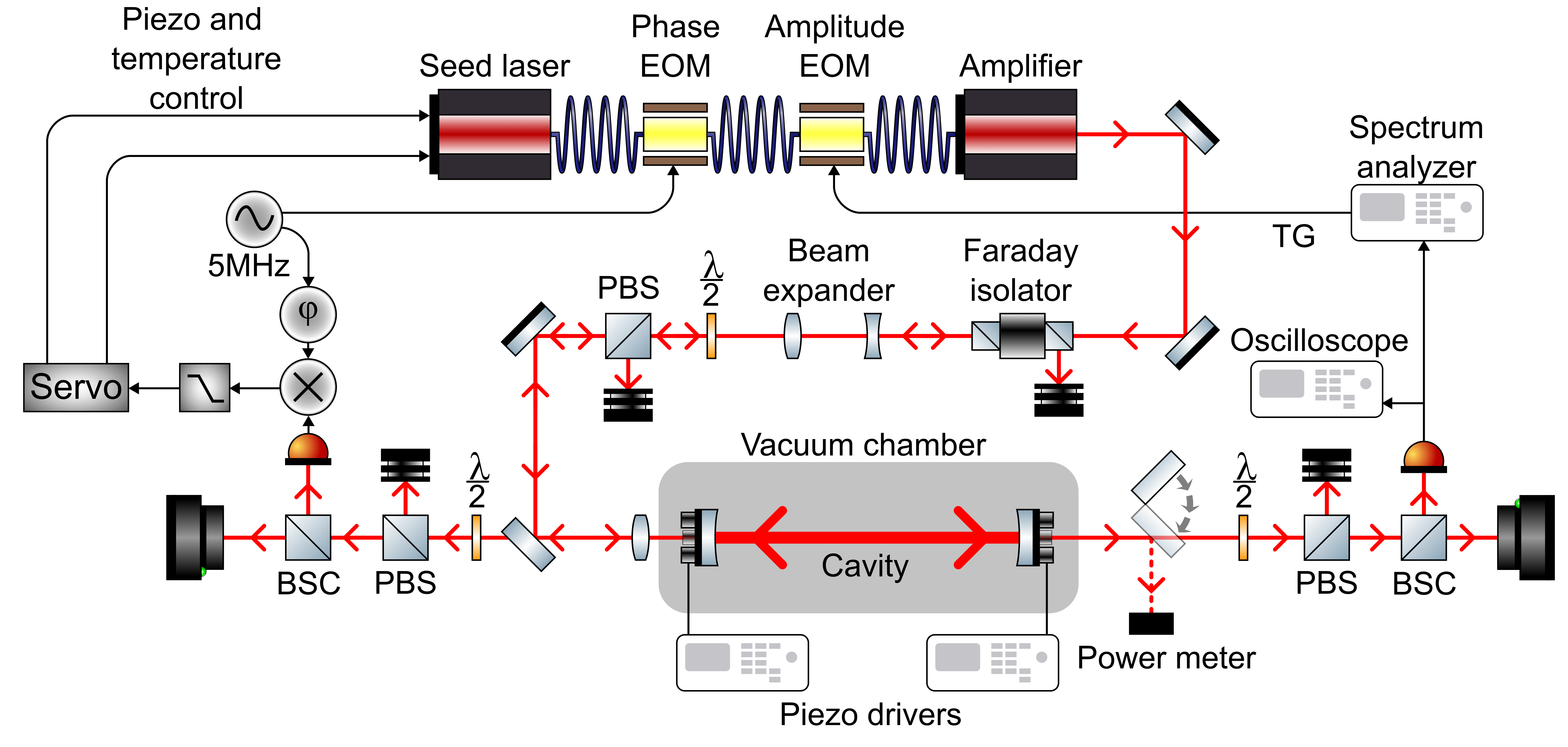}
    \caption{\textbf{Schematic of the high-power resonator.} A near-concentric resonator is suspended in a vacuum chamber. The input beam, generated by a low-noise seed laser, is phase-modulated by a fiber electro-optical modulator (EOM) and amplitude-modulated by a second EOM. The beam is then amplified in a fiber amplifier and, after passing through a Faraday isolator, is directed into a mode-matching beam expander. It is then attenuated to the desired power and directed into the resonator. A portion of the reflected beam is used for PDH locking. A spectrum analyzer with a tracking generator (TG) is used to modulate the amplitude of the input beam at a frequency close to the free spectral range of the resonator and to record the modulation of the transmitted light, to continuously monitor the resonator line shape at high power.  Notation: $\lambda/2$ -- half-wave plate, PBS -- polarizing beamsplitter cube, BSC -- beamsplitter cube.}
    \label{fig:Optical scheme}
\end{figure}

A schematic of our experimental setup is shown in Fig.~\ref{fig:Optical scheme}. A symmetric near-concentric optical resonator is suspended in a vacuum chamber. The mirror radii of curvature are $0.5$ m and the cavity length is $0.99$ m, corresponding to a NA of $0.0026$ at a wavelength of $1064$ nm. The values of the cavity length and the NA are confirmed by measurement of FSR and the spectrum of the transverse cavity modes. The resonator frame consists of four fused silica rods connected by stainless steel rings. The low thermal expansion coefficient and low absorption of the rods minimize the cavity length changes caused by thermal expansion due to absorption of the light scattered in the cavity. The tilt of each mirror is adjusted using two micrometric screws and three piezo actuators. 

The 120 W input laser beam at a wavelength of 1064 nm was generated by a master oscillator-power amplifier (MOPA) system comprising a non-planar ring oscillator seed laser (Coherent Mephisto) and an ytterbium-doped fiber amplifier (Azurlight).  The laser frequency was locked to the cavity resonance using the Pound-Drever-Hall scheme.

\begin{figure}
    \centering
    \includegraphics[width=8cm]{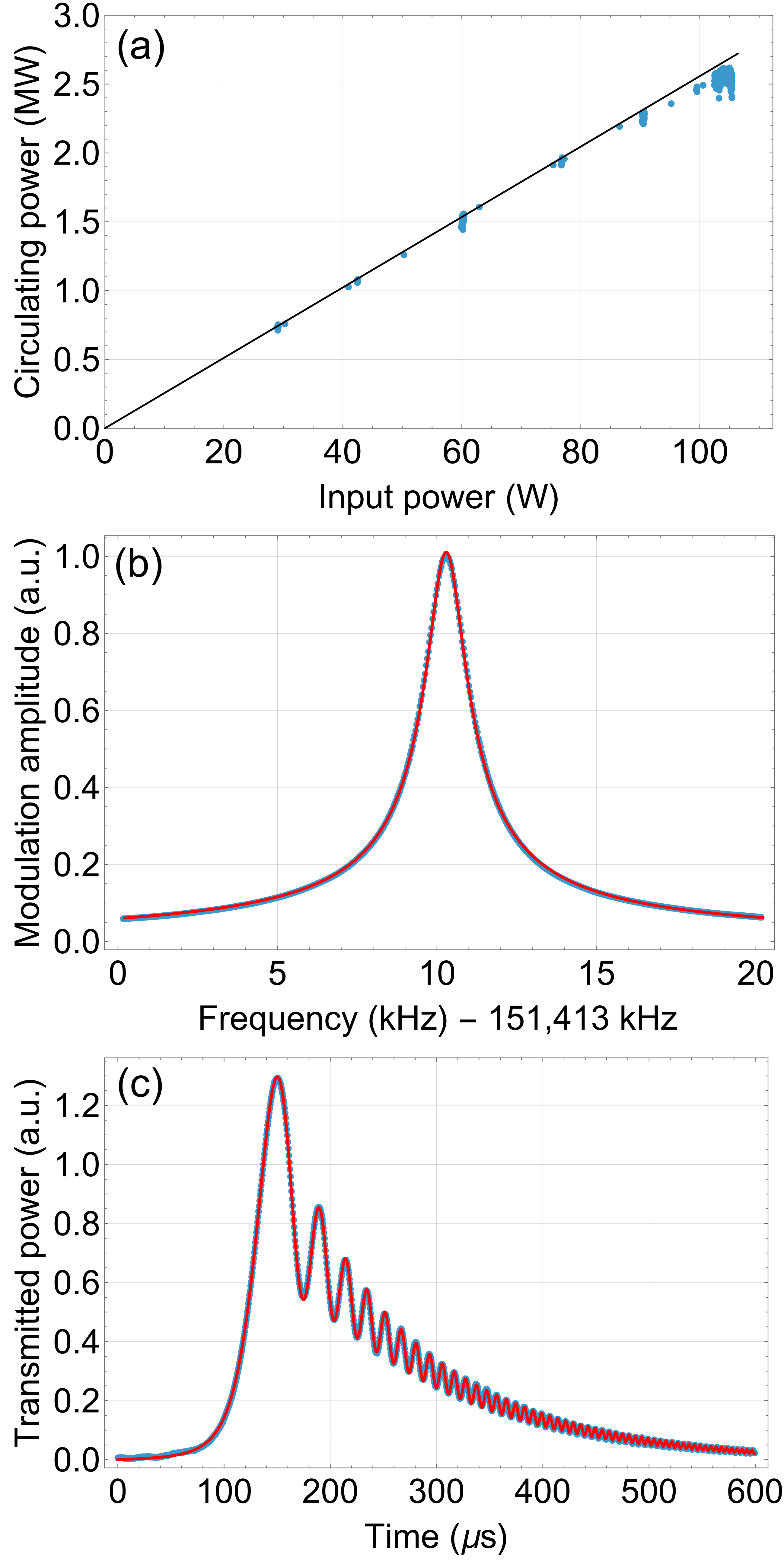}
    \caption{(a) Circulating intracavity power as a function of the input power, calculated by dividing the transmitted power by the transmission coefficient of the output mirror. The black line corresponds to a coupling coefficient of $0.72$, cavity mirror transmission $T = 23.1$ ppm, and reflection loss of $L=2.3$ ppm. (b) The experimentally measured cavity resonance shape (blue dots) and the Lorentzian fit (red line) as a function of modulation frequency. (c) Cavity ring-down signal. Blue dots indicate the experimental data, and the red line represents the fitted model curve.}
    \label{fig:Experiment}
\end{figure}

Absorption of light in the mirror coatings results in a heat load of a few watts on the mirror surface, which can elevate the mirror temperature above the operating range of the dielectric coatings. To overcome this limitation, we chose sapphire as the substrate material and implemented a low-vibration cooling system that stabilized the mirror temperature at -50$\degree$C. The mirrors were cooled by copper braids connecting them to a cold plate cooled by liquid nitrogen. The high thermal conductivity of sapphire ($70 \mbox{ W}/ (\mbox{m} \cdot \mbox{K})$ \cite{touloukian_thermophysical_1971}) ensures that the temperature variation across the mirror does not exceed a few degrees. 

The circulating power in the resonator was measured by monitoring the power transmitted through it, divided by the transmission coefficient of the resonator mirror. The laser frequency was initially locked to the cavity resonance at an input power of  29~W. The input power was then gradually increased to approximately $105$ W, at which point the system operated steadily. The observed dependence of the circulating power on the input power is shown in Fig.~\ref{fig:Experiment}a. The maximum circulating power achieved was $2.5$ MW.

The accuracy of the circulating power measurement depends on the resonator mirror transmission coefficients remaining unchanged at high power. We verified this by applying a weak amplitude modulation to the input laser beam and measuring the modulation amplitude of the transmitted beam. Scanning the modulation frequency over a range of a few kHz around the frequency $\Delta_\text{FSR}$, we measured the resonator line shape (see Fig.~\ref{fig:Experiment}b) and extracted the linewidth every 2 seconds. No change in the resonator linewidth was observed at high circulating power compared to low-power operation. 

The linewidth measured by scanning the modulation frequency was $611 \pm 18$ Hz, in good agreement with the cavity ring-down time measurement shown in Fig. \ref{fig:Experiment}c, corresponding to a cavity linewidth of  $618 \pm 17$ Hz. The ring-down measurement was performed by sweeping the laser frequency across the cavity resonance and measuring the transmitted signal as a function of time. The signal was fitted to a model using only the sweep rate, cavity linewidth, and overall signal amplitude as fitting parameters \cite{schwartz_near-concentric_2017}. %
The power transmission coefficient of the output mirror, measured prior to resonator assembly, was $T =23.1$ ppm, corresponding to an average loss coefficient per mirror of $2.3\pm0.8$ ppm.

We have thus shown that circulating powers in the megawatt range can be achieved in a near-concentric resonator.

\section{Discussion}
\subsection{Resonant enhancement of the signal sidebands} \noindent
We note that the nonlinear-optical signal in our scheme arises not only from the resonant enhancement of the input fields, but also from the coherent accumulation of the sideband amplitude over the lifetime of the signal photons in the resonator. In our experimentally realized resonator, this corresponds to an intracavity signal sideband amplitude enhancement of $F / \pi\simeq 4\cdot10^4$. 

The physical mechanism of the signal enhancement in the cavity can be described as follows: the pump and probe fields generate nonlinear electric ($\textbf{P}$) and magnetic ($\textbf{M}$) polarizations in the overlapping focal volumes of the resonators. The polarization vectors oscillate in time, emitting light at combinations of input frequencies. In free space, that emission would be far too small to detect, generating less than a single photon over the duration of the experiment. However, due to the resonant enhancement of emission into the cavity mode, the polarization generates a measurable signal photon flux. 

\subsection{Residual gas nonlinearity}\noindent
Measurement of EMNV can be hampered by the nonlinear response of the residual gas in the vacuum chamber. The critical pressure at which the four-wave mixing in gas equals that due to EMNV is approximately $5\cdot10^{-12}$~mbar \cite{robertson_experiment_2021}. Such low pressure can be achieved in a thoroughly baked-out vacuum chamber \cite{stoltzel_comparison_2023}, but the use of piezo actuators for fine-tuning of the mirror angles complicates heating the chamber above the depolarization temperature of the piezo stacks. However, a differentially pumped low-outgassing sub-chamber can be used to achieve XHV pressures in the interaction region.

\subsection{Further improvements of the SNR}\label{Possible upgrades}\noindent
The SNR of the EMNV detection scheme presented above can be increased by implementing one or more of the following upgrades. 

First, the probe light exiting the resonator can be recycled by a frequency-selective reflector that returns most of the probe power to the resonator, while transmitting the sideband light toward the detector. This can be accomplished, e.g., by a Michelson interferometer with a path difference between its arms tuned to equal the resonator length. 
Likewise, a Michelson interferometer placed before the resonator input mirror can transmit the probe light into the cavity while reflecting the sideband light leaking through the input mirror back into the resonator. Thus, both the signal and probe light can be recycled, reducing the required integration time by a factor of four.

In addition, the pump resonator can be configured to maximize circulating power at a given loss by using an output mirror with near-zero transmission and matching the input mirror transmission to the resonator round-trip loss. This can double the circulating pump power and thus provide another fourfold reduction of the measurement time.

\begin{figure}
    \centering
    \includegraphics[width=8cm]{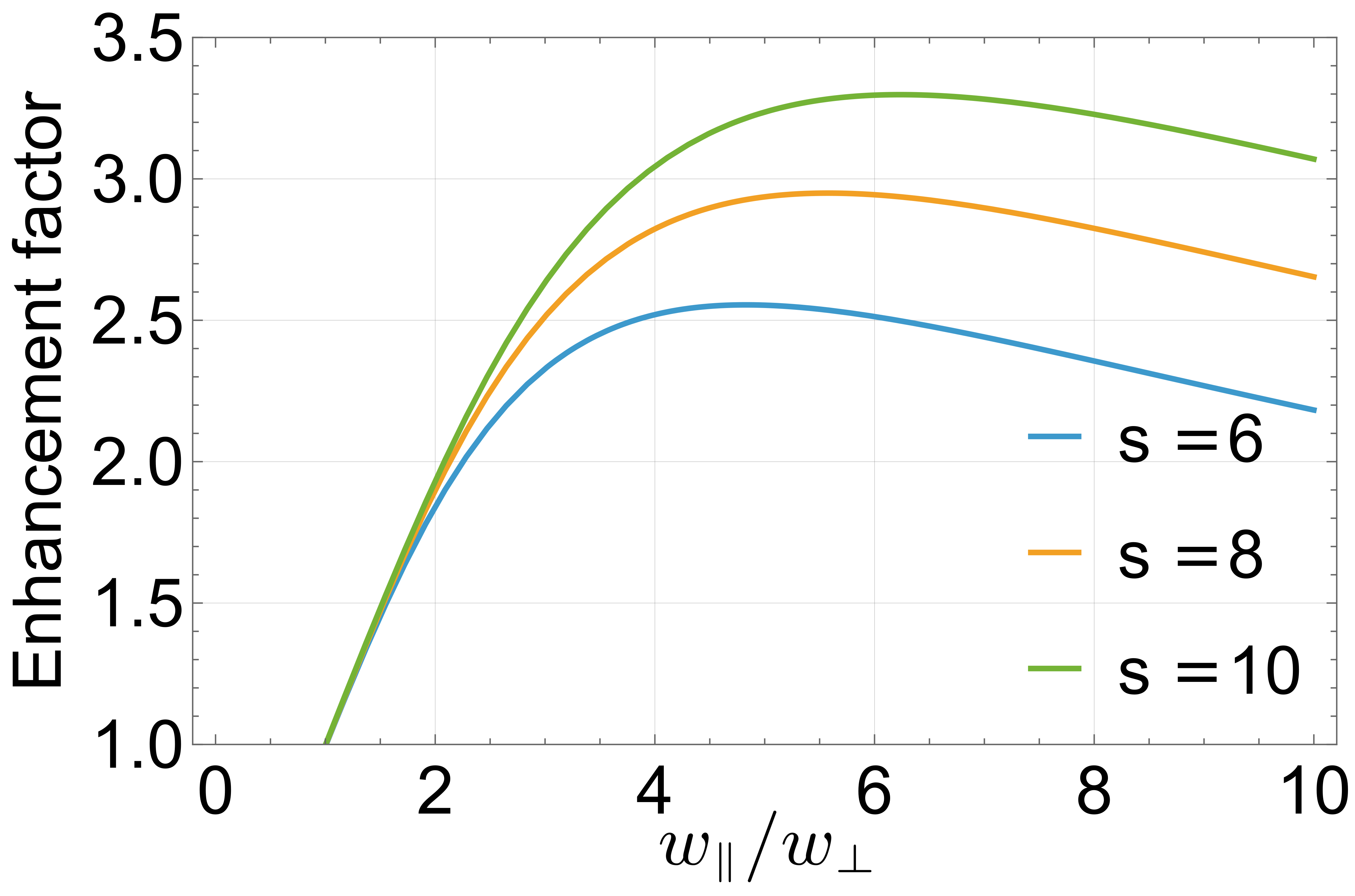}
    \caption{The sideband amplitude enhancement factor $\kappa$ as a function of the mode aspect ratio $w_{\scriptscriptstyle \parallel}/w_{\scriptscriptstyle \perp}$,  plotted for several values of the separation parameter $s$.}
    \label{fig:Astig}
\end{figure}

Another way to increase the SNR is to use resonators with astigmatic mirrors supporting fundamental modes with an optimized aspect ratio. At a given separation factor $s$, shrinking the mode waist in the direction orthogonal to the resonator crossing plane can increase the focal intensity, at the expense of a slight reduction of the interaction length.  %
At a small angle between resonators and a separation factor $s\gg 1$, the sideband amplitude enhancement factor $\kappa$ due to mode astigmatism takes the form:
\begin{equation}
\label{astig}
    \kappa = \sqrt{\frac{s}{\pi}} f\left(\frac{w_{\scriptscriptstyle \parallel}^2}{s\, w_{\scriptscriptstyle \perp}^2}\right),
\end{equation}
where $w_{\scriptscriptstyle \parallel}$ and $w_{\scriptscriptstyle \perp}$ are, respectively, the in-plane and out-of-plane mode waists, and $f(p) = \sqrt{p} \int \limits_{-\infty}^{\infty} \frac{\exp (-q^2) }{\sqrt{1+p^2 q^2}}  dq$. To simplify the expression, this formula is derived for the case of identical wavelengths of the pump and probe light. The enhancement factor $\kappa$ as a function of the waist ratio $w_{\scriptscriptstyle \parallel}/w_{\scriptscriptstyle \perp}$ is shown in Fig.~\ref{fig:Astig} for several values of the separation parameter $s$. At our preferred separation of $s=8$, the maximum value of the enhancement factor is $\kappa = 2.9$, corresponding to about eightfold reduction of the measurement time.

A further large improvement of the SNR can be achieved by exciting multiple modes in the pump and probe resonators, so that the pump and probe fields form synchronized pulses inside the resonators \cite{bullis_interferometric_2025}. At a given average circulating power in the resonators, the signal amplitude scales inversely with the pulse length until it becomes shorter than the length of the intersection of the two beams. For the parameters listed above, that corresponds to a pulse duration of approximately $30$ ps and a spectral width of about $15 \;\mbox{GHz} \simeq 100 \, \Delta_\text{FSR}$, resulting in a measurement time reduction factor of approximately 1000.    

Finally, squeezed light has been used to increase the SNR in high-sensitivity interferometric measurements \cite{jia_squeezing_2024}. In our case, the high circulating power could facilitate the use of intra-cavity nonlinearity to generate squeezed states of light in the probe resonator. The nonlinearity can be controlled by engineering the nonlinear-optical properties of the cavity mirror materials, tailoring their opto-mechanical response, or introducing an additional squeezing device into the cavity.  Strong intra-cavity nonlinearity can be used, e.g., to generate squeezed vacuum in the sideband modes, reducing shot noise in the heterodyne measurements. 

\subsection{Summary} \noindent
We proposed a method for detecting EMNV based on cross-phase modulation of CW laser beams resonantly enhanced in high-finesse resonators. We showed theoretically that it can generate optical signals strong enough to be detected via heterodyne measurements. Furthermore, we experimentally demonstrated an optical resonator reaching a circulating power of $2.5$ MW, within a factor of two of the power needed to measure EMNV at the level predicted by the Euler-Heisenberg effective Lagrangian. These results establish a realistic path to a tabletop all-optical measurement of EMNV.
\section{Acknowledgments}
\noindent
We thank Holger Mueller, Jeremy J. Axelrod, Gilad Perez, and Noam Tal Hod for insightful conversations. We gratefully acknowledge financial support from the Minerva Stiftung (Project No. 714444), the Leona M. and Harry B. Helmsley Charitable Trust, and the Zuckerman STEM Leadership Program. 

\bibliographystyle{opticajnl}
\bibliography{references}

\end{document}